\documentclass[reprint,superscriptaddress,prb,floatfix]{revtex4-2}

\usepackage[T1]{fontenc}
\usepackage{graphicx}
\usepackage{dcolumn}
\usepackage{siunitx}
\DeclareSIUnit{\barpressure}{bar}
\DeclareSIUnit{\langmuir}{L}
\DeclareSIUnit\angstrom{\protect \text {Å}}
\DeclareSIUnit{\mubpercr}{\protect $\mu_B$/\text{Cr}}
\usepackage{amsmath}
\usepackage{xcolor}
\usepackage{comment}
\usepackage{physics}
\usepackage{ulem}
\DeclareGraphicsExtensions{.pdf,.png}

\begin{document}
\preprint{APS/123-QED}

\title{Orbital mixing and strong Hund's coupling stabilize spin order in van der Waals ferromagnet CrI\textsubscript{3}}

\author{A. De Vita}
\email{alessandro.de.vita@tu-berlin.de}
\affiliation{Institut für Physik und Astronomie, Technische Universität Berlin, Straße des 17 Juni 135, 10623 Berlin, Germany\looseness=-1}%
\affiliation{Fritz Haber Institute of the Max Planck Society, Faradayweg 4--6, 14195 Berlin, Germany\looseness=-1}%
\author{S. Stavri\'c}
\email{stavric@vin.bg.ac.rs}
\affiliation{Vin\v{c}a Institute of Nuclear Sciences - National Institute of the Republic of Serbia, University of Belgrade, P. O. Box 522, RS-11001 Belgrade, Serbia\looseness=-1}%
\author{R. Sant}%
\affiliation{ESRF, The European Synchrotron, 71 Avenue des Martyrs, CS40220, 38043 Grenoble Cedex 9, France\looseness=-1}%
\affiliation{Dipartimento di Fisica, Politecnico di Milano, Piazza Leonardo da Vinci 32, I-20133 Milano, Italy\looseness=-1}%
\author{N. B. Brookes}
\affiliation{ESRF, The European Synchrotron, 71 Avenue des Martyrs, CS40220, 38043 Grenoble Cedex 9, France\looseness=-1}%
\author{I. Vobornik}
\affiliation{Consiglio Nazionale delle Ricerche CNR-IOM, Unità di Trieste, Strada Statale 14, km 163.5, 34149 Basovizza (TS), Italy\looseness=-1}%
\author{G. Panaccione}
\affiliation{Consiglio Nazionale delle Ricerche CNR-IOM, Unità di Trieste, Strada Statale 14, km 163.5, 34149 Basovizza (TS), Italy\looseness=-1}%
\author{S. Picozzi}
\affiliation{Department of Materials Science, University of Milan-Bicocca, Via Roberto Cozzi 55, 20125 Milan, Italy\looseness=-1}%
\affiliation{Consiglio Nazionale delle Ricerche CNR-SPIN, c/o Università degli Studi ‘G. D’Annunzio’, 66100 Chieti, Italy\looseness=-1}
\author{M. Wolf}
\affiliation{Fritz Haber Institute of the Max Planck Society, Faradayweg 4--6, 14195 Berlin, Germany\looseness=-1}%
\author{L. Rettig}
\affiliation{Fritz Haber Institute of the Max Planck Society, Faradayweg 4--6, 14195 Berlin, Germany\looseness=-1}%
\author{R. Ernstorfer}
\email{ernstorfer@tu-berlin.de}
\affiliation{Institut für Physik und Astronomie, Technische Universität Berlin, Straße des 17 Juni 135, 10623 Berlin, Germany\looseness=-1}%
\affiliation{Fritz Haber Institute of the Max Planck Society, Faradayweg 4--6, 14195 Berlin, Germany\looseness=-1}%
\author{T. Pincelli}
\email{pincelli@tu-berlin.de}
\affiliation{Institut für Physik und Astronomie, Technische Universität Berlin, Straße des 17 Juni 135, 10623 Berlin, Germany\looseness=-1}%
\affiliation{Fritz Haber Institute of the Max Planck Society, Faradayweg 4--6, 14195 Berlin, Germany\looseness=-1}%


\date{\today}

\begin{abstract}

Recent years have seen a vast increase in research into van der Waals magnetic materials. In many of these systems, magnetism is introduced via light 3\textit{d}-transition metal elements, combined with chalcogenides or halogens. Despite the high technological promise in the field of spintronics, the connection between the \textit{d}-orbital configuration and the occurrence of low-dimensional magnetic order is currently unclear. Here we address the prototypical two-dimensional ferromagnet CrI\textsubscript{3}, via complementary spectroscopies and density functional theory calculations. We reveal the electronic structure and orbital character of bulk CrI\textsubscript{3} in the paramagnetic and ferromagnetic phases, describing the couplings underpinning its energy diagram, and providing a robust experimental demonstration that the stabilization of ferromagnetism is attributable to orbital mixing between I \textit{p} and Cr \textit{e\textsubscript{g}} states, and to the presence of strong Hund's coupling. These findings reveal the microscopic connection between orbital and spin degrees of freedom, providing fundamental insights into the behavior of low-dimensional magnetic materials.

\end{abstract}


\maketitle

\section{Introduction}

Van der Waals (vdW) magnets constitute a versatile platform where exotic quantum states can be realized in view of applications in spin- and orbitronics \cite{Samarth2017,Gibertini2019,Kajale2024}. In particular, much interest has been drawn to the orbital degree of freedom in quantum materials \cite{Hoffmann2015,Go2018,Vedmedenko2020}. Given the dimensional confinement in vdW layers, a working approximation of localized, orbitally-pure electronic states would seem adequate. On the other hand, orbital-specific correlations and couplings are significant to solve ground state and collective excitations in these systems \cite{Stavropoulos2019,He2025,Occhialini2025}. Understanding the interplay between orbital occupation and electronic ordering, as well as the \textit{d}-electron contribution to the emergence of low-dimensionality magnetism, represents a fundamental research challenge.

Among vdW magnets, the layered ferromagnet CrI\textsubscript{3} exhibits long-range magnetism down to the monolayer limit induced by magnetic anisotropy, with stacking-dependent magnetic order and a pronounced magnetoelectric response \cite{Gong2017,Huang2017,Fernandez2020,Jiang2018a,Huang2018,Jiang2018b,Jiang2019,Soriano2020}. This makes CrI\textsubscript{3} an ideal candidate to probe the influence of orbital phenomena on the magnetic configuration. In particular, Kim \textit{et al.} \cite{Kim2019} have proposed that the stabilization of magnetic interactions in CrI\textsubscript{3} involves a strong contribution originating from \textit{p-d} covalency. While this proposition has been well-received, few spectroscopic studies have been aimed at providing a detailed experimental investigation, mostly limited to the non-magnetic phase of CrI\textsubscript{3} due to its extreme hygroscopicity, sensitivity to contamination, and charging at low temperatures \cite{Kundu2020,DeVita2022}.

In this work, we present a multifaceted study of the electronic structure of CrI\textsubscript{3} above and below the Curie temperature $T_C=\SI{61}{\K}$, combining Angle-Resolved Photoemission Spectroscopy (ARPES) and X-ray Magnetic Circular Dichroism (XMCD) with density functional theory (DFT) and multiplet cluster calculations. Absorption data demonstrate that CrI\textsubscript{3} displays a covalent character caused by orbital mixing between Cr \textit{e\textsubscript{g}} and I 5\textit{p} states. Electronic band structure modifications attributed to exchange-induced splitting and Hund's energy gain were detected by temperature-dependent ARPES and supported by DFT calculations. The combined techniques bring robust experimental proof that the Cr-I hybridization is responsible for the stabilization of ferromagnetism in CrI\textsubscript{3}, and establishes the role of Hund's coupling as a major driving force in this system. Our findings provide essential information on the electronic structure of this material, and offer new insight on the microscopic orbital mechanisms underpinning its properties. 

\section{Methods}

\textbf{Photoelectron Spectroscopy} -- ARPES spectra have been acquired with XUV \textit{p}-polarized photons at \SI{21.7}{\eV}, provided by the high-harmonics generation (HHG) setup in the ARPES laboratory at the Fritz Haber Institute of the Max Planck Society, Berlin \cite{Puppin2019,Maklar2020}. The overall energy resolution has been estimated to be $\approx\SI{150}{\milli\eV}$; the spot size is \SI{80}{\micro\metre} FWHM. Commercially available CrI\textsubscript{3} crystals have been mounted on sample holders while in N\textsubscript{2} atmosphere and low, red light conditions using silver epoxy glue (HD20E, Epotek), and cleaved in UHV ($p<\SI{8e-11}{\milli\barpressure}$) before the measurement. To compensate for sample charging at low temperatures, a \SI{400}{\nano\metre}, CW laser has been used in conjunction with XUV during the measurement. Like the similar halide VI\textsubscript{3} \cite{DeVita2026}, CrI\textsubscript{3} displays pronounced light sensitivity under beam: for this reason, we monitored the peaks at \SI{-1.3}{\eV} and \SI{-2}{\eV} as proxies to spot variations in the lineshape as a function of measurement time. Since the beam footprint on the sample surface is roughly \qtyproduct{50x100}{\um}, any beam damage is a local effect, and moving to another spot generally recovers the initial lineshape. Compared to VI\textsubscript{3}, CrI\textsubscript{3} is somewhat more sensitive to irradiation, therefore surface degradation from adsorbates cannot be effectively countered by increasing the photon flux. (\textit{E}, \textit{k}) spectra have been collected with a TOF momentum microscope at \ang{65} incidence angle and normal emission conditions. 

Resonant-ARPES (ResPES) measurements at resonant photon energy have been acquired at the APE-LE beamline of the Elettra Synchrotron Radiation Facility \cite{Panaccione2009}. CrI\textsubscript{3} crystals have been mounted on sample holders while in N\textsubscript{2} atmosphere using silver epoxy glue (HD20E, Epotek), and cleaved in a pressure better than $\SI{5e-9}{\milli\barpressure}$ before the measurement. Samples were aligned with the slit along the $\bar{\Gamma}-\bar{\mathrm{K}}$ direction, and kept at \ang{45} incidence angle and normal emission conditions. 

The energy axis in all the (\textit{E}, \textit{k}) spectra has been aligned such that the valence band maximum (VBM) corresponds to zero energy. We employed a standard approach in semiconductor physics, as described in Refs. \cite{Kraut1980,Katnani1983}: we linearly extrapolate the spectral leading edge to the baseline, and take the intersection as a proxy for VBM. An example EDC is shown in Supplemental Material, Fig.~S1.

\textbf{Absorption Spectroscopy} -- XAS and XMCD measurements have been acquired at beamline ID32 of the European Synchrotron Research Facility (ESRF) \cite{Kummer2016,Brookes2018}. Commercially available CrI\textsubscript{3} crystals stored in Ar atmosphere ($<0.5$ ppm O\textsubscript{2}, $<0.5$ ppm H\textsubscript{2}O) have been transferred in inert static atmosphere and cleaved under N\textsubscript{2} flow inside the loadlock chamber, to expose the (0001) crystallographic plane of the clean surface avoiding oxidation of the highly hygroscopic surface due to moisture; the sample was held in ultra-high vacuum (UHV, pressure $<\SI{3e-10}{\milli\barpressure}$). X-ray absorption spectroscopy (XAS) spectra have been acquired in total electron yield at normal incidence and normalized by the intensity collected on a Au mesh in front of the sample stage. The beam has almost \SI{100}{\percent} degree of linear and circular polarization, and the setup has resolving power better than 5000. A  Cr\textsubscript{2}O\textsubscript{3} calibration sample located after the exit slit \cite{Brookes2018} has been used as reference for the energy scale. XMCD spectra are the difference between left- and right- circularly polarized light spectra, measured at remanence after zero-field cooling from room temperature to \SI{25}{\K} following application of \SI{0.5}{\tesla} out-of-plane magnetic field. Measurements have been performed at \SI{30}{\K} and \SI{90}{\K}. The beam spot diameter at sample position at normal incidence is \qtyproduct{100x100}{\um}.

\textbf{DFT calculations} -- DFT calculations were performed by using the Vienna {\it Ab-initio} Simulation Package ({\sc VASP})~\cite{Kresse96,Kresse96b,Kresse1999Jan}. The lattice constant of $a=\SI{6.817}{\angstrom}$ and a Cr-I bond length of $\SI{2.689}{\angstrom}$ were determined by performing the spin-polarized relaxation of the monolayer structure in the ferromagnetic state using the PBEsol exchange-correlation functional \cite{Perdew2008Apr}. A plane wave energy cutoff of \SI{350}{\eV} was used, and the Brillouin zone (BZ) was sampled with a 24 $\times$ 24 $\times$ 1 \textit{k}-point grid. To treat the Cr 3\textit{d} states, we adopted the DFT+\textit{U} approach as introduced by Liechtenstein \textit{et al.} \cite{Liechtenstein1995}, and tested several Hubbard \textit{U} and Hund \textit{J} values were tested to find the best agreement between with experiments (details on particular choice of $U$ and $J$ parameters are given below). Since the I $5p$-dominated VBM states are important to our analysis and given iodine's strong SOC effects (note that $Z({\rm I})=53$), we included SOC with non-collinear DFT to properly describe these states. Notably, including SOC has minimal influence on the energy positions of the deeper Cr $3d$ states.

\begin{figure*}[htb]
\includegraphics[width=0.9\linewidth]{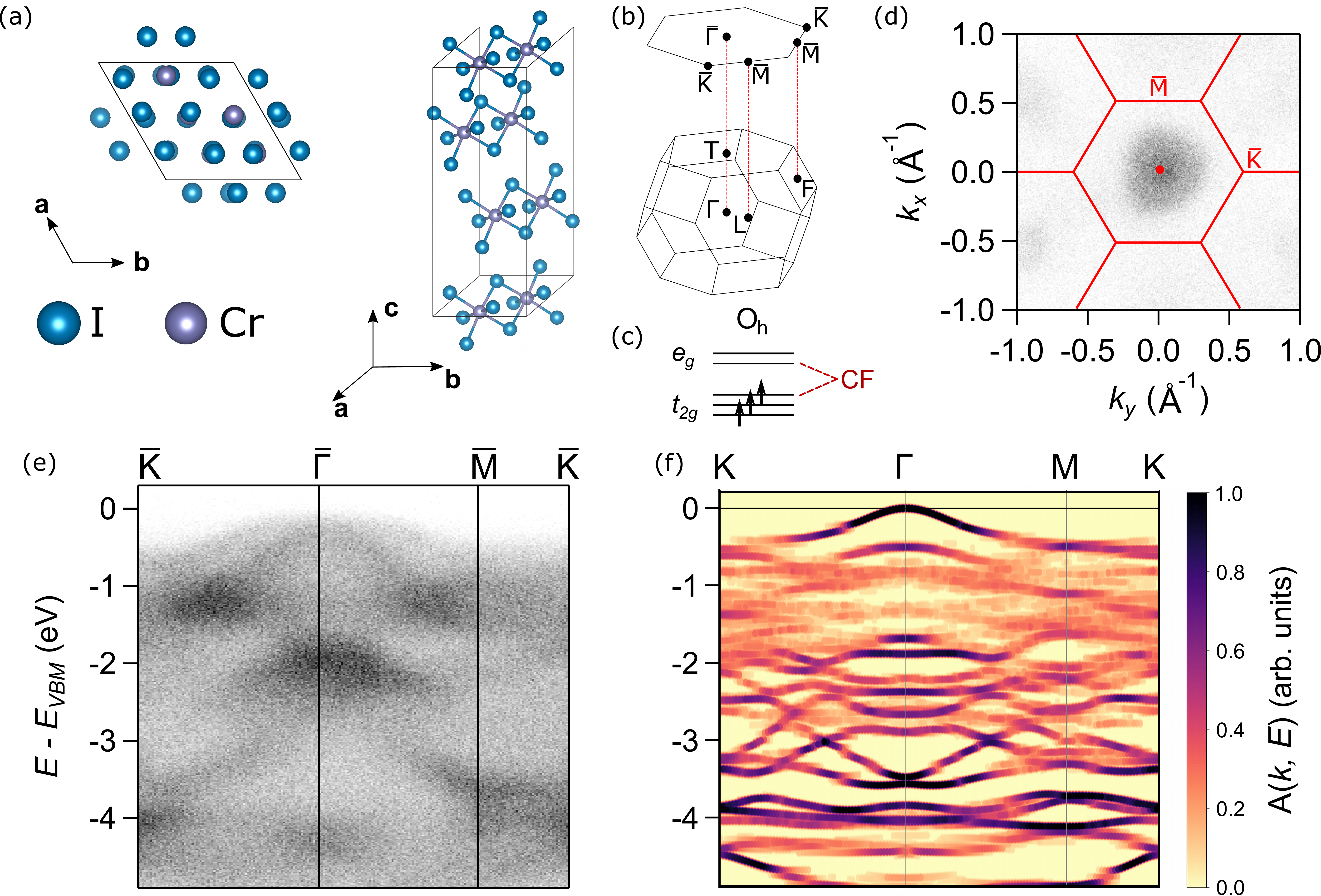}
\caption{\protect\label{fig:Fig1} (a) Top and side hard sphere view of CrI\textsubscript{3} crystal structure, where each Cr centre is coordinated by six I atoms. (b) Bulk first Brillouin zone of CrI\textsubscript{3} and corresponding surface projection; high-symmetry points are indicated in both cases. (c) Crystal field splitting and nominal electron filling for O\textsubscript{h} symmetry. (d) Photoemission intensity (greyscale, darker represents higher electron count) of the isoenergetic ($k_x$, $k_y$) surface at $E-E_{VBM}=\SI{-0.1}{\eV}$, at $h\nu=\SI{21.7}{\eV}$ photon energy and $T=\SI{300}{K}$. The red overlay highlights high-symmetry points and lines. (e) ARPES (\textit{E}, \textit{k}) spectrum of  CrI\textsubscript{3}  valence band dispersion along high-symmetry lines in the same experimental conditions as (d). (f) The spectral function $A({\bf k}, E)$ of the paramagnetic state calculated with DFT.
}
\end{figure*}

\section{Main text}

The (low-temperature) rhombohedral ($\mathrm{R\bar{3}}$) crystal structure \cite{Momma2011}, bulk and surface Brillouin zone (BZ), as well as the nominal Cr 3\textit{d} orbital filling are depicted in Fig.~\ref{fig:Fig1}a-c. Firstly, we consider ARPES data obtained on CrI\textsubscript{3} valence band (VB) at $T=\SI{300}{\K}$. The isoenergetic surface and the (\textit{E}, \textit{k}) spectra along the high-symmetry directions in the surface BZ are presented in Fig. \ref{fig:Fig1}d-e. Measurements are consistent with previous results \cite{DeVita2022}: the VBM is dominated by strongly dispersive states, displaying a slight trigonal warping (Fig. \ref{fig:Fig1}d), whereas at higher binding energies (BEs) from \SIrange{-1}{-2}{\eV} the electronic states appear more localized.

We model the CrI\textsubscript{3} electronic structure using non-collinear DFT with spin-orbit coupling (SOC) included. Due to the vdW nature of bulk CrI\textsubscript{3} and the presence of SOC-induced anisotropy, the band structure evolution across the magnetic phase transition is dominated by intralayer interactions, with interlayer effects playing a secondary role. We therefore model CrI\textsubscript{3} as a monolayer. The paramagnetic (PM) state is reproduced by using the $3 \times 3 \times 1$ supercell and introducing magnetic disorder through the special quasi-random structures (SQS) method \cite{Zunger1990Jul,Angqvist2019Jul} (see the Appendix for technical details). To reveal how disordering of Cr magnetic moments affects the electronic bands across different energy ranges, we compute the spectral function $A({\bf k}, E)$ and unfold it to the unit cell \cite{Dirnberger2021Jun} to enable direct comparison with ARPES data.  
%
We systematically evaluate various Hubbard \textit{U} and Hund \textit{J} parameters for the Cr 3\textit{d} orbitals, ultimately determining that $U=\SI{1.3}{\eV}$ and $J=\SI{0.9}{\eV}$ yield optimal agreement of the calculated spectral function with ARPES measurements in terms of band dispersion at the VBM, high-intensity pockets between \SIrange{-1}{-2}{\eV}, and other features at higher BEs (Fig.~\ref{fig:Fig1}e-f). We note that the $(U, J)$ parameters were calibrated by tracking the energy position of the occupied $\text{Cr}\ 3d$ band maxima rather than the band gap magnitude, which is often the standard metric in combined experimental-computational studies. This approach is necessitated by the insensitivity of the $\text{CrI}_3$ band gap to the Hubbard $U$ and $J$ values. Specifically, the valence band maximum consists primarily of $\text{I}\ 5p$ states, while the conduction band minimum is composed of spin-majority $\text{Cr}\ 3d$ states, both of which are largely unaffected by the local Hubbard interaction on the Cr sites. Notwithstanding this insensitivity, our selected $(U, J)$ values yield a DFT band gap of \SI{1.22}{\eV} in the FM state and \SI{1.27}{\eV} in the PM state, values that are in excellent agreement with previously reported calculations and experimental optical gap of $\sim \SI{1.2}{\eV}$ \cite{Marfoua2020,Nguyen2021,Dillon1965,Jin2020}. Spectral functions calculated with different $(U,J)$ parameters are presented in Fig.~S6. Flat bands around $\Gamma$ appearing at \SI{-2}{\eV} in Fig.~\ref{fig:Fig1}e are well captured by DFT, as well as the bundle of bands at \SI{-4}{\eV} spreading throughout the BZ. While the energy positions of bands can be compared between ARPES measurements and DFT-calculated spectral functions, their intensities can be only partly related, because the DFT-calculated spectral function does not reproduce the intensity nuance of ARPES signal arising from many-body interactions and Bloch wavefunction symmetries.

\begin{figure}[h]
\includegraphics[width=0.95\linewidth]{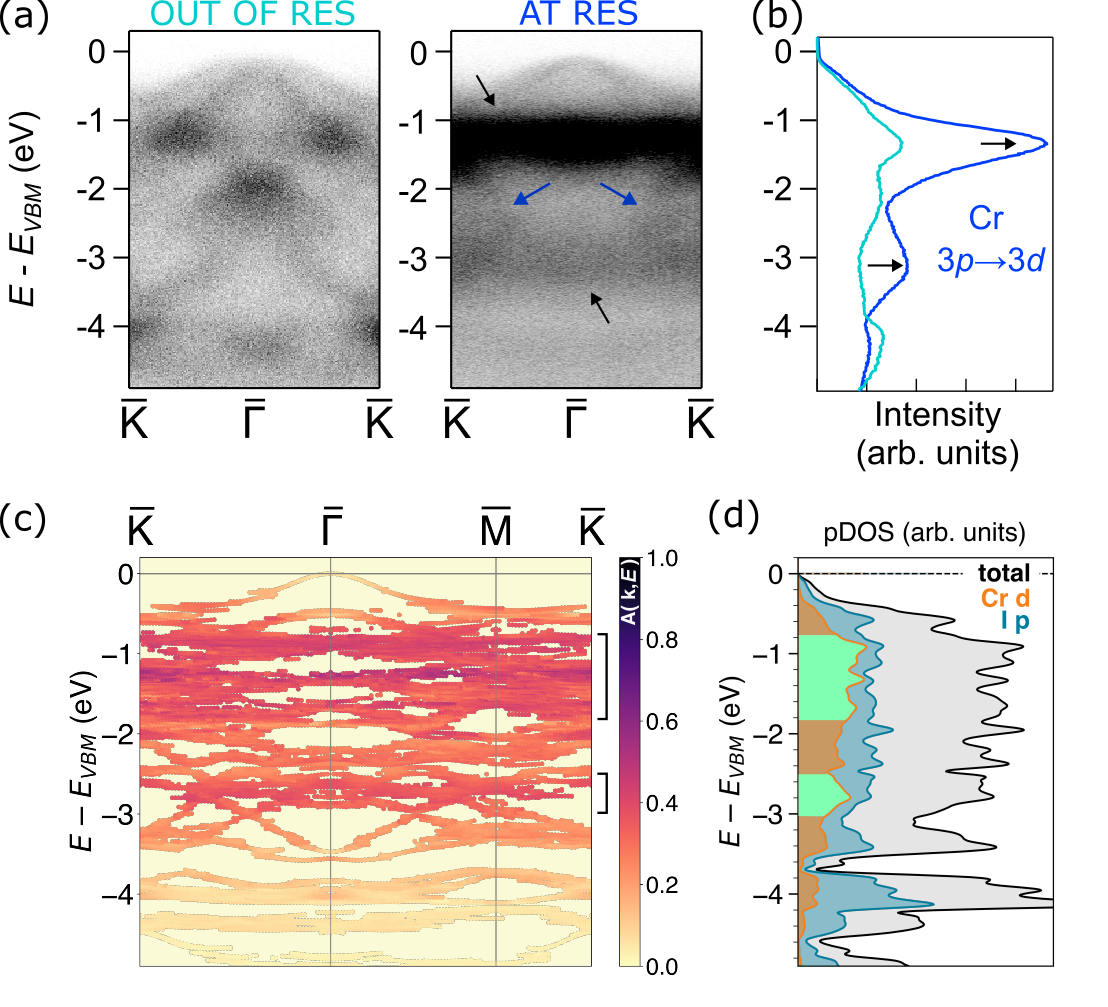}
\centering
\caption{\protect\label{fig:Fig3}(a) ARPES (\textit{E}, \textit{k}) spectra along the $\bar{\mathrm{K}}-\bar{\Gamma}-\bar{\mathrm{K}}$ direction, at \SI{21.7}{\eV} (\textit{left}, out of resonance) and \SI{49}{\eV} (\textit{right}, at resonance), at \SI{300}{\K}. Black and blue arrows point at the features with increased spectral weight. (b) $k$-integrated EDCs of the spectra in (a): out of resonance is teal, at resonance is blue. Black arrows mark the two prominent increases in intensity. (c) Cr 3\textit{d} bands  calculated with DFT in the PM state. (d) Atomic-orbital projected density of states in the PM state. Areas shaded in green highlight the local maxima of Cr 3\textit{d} states, as described in the main text.}
\end{figure}

To further assess how the charge distribution affects the VB, we acquired ARPES spectra out-of- and at-resonance with the Cr 3\textit{p} core level, displayed in Fig. \ref{fig:Fig3}a.
At resonance, the non-dispersive feature centered at $\sim\SI{-1.3}{\eV}$ is strongly enhanced, as emphasized also by the corresponding energy distribution curves (EDCs) in Fig. \ref{fig:Fig3}b. We attribute this manifold to states with Cr 3\textit{d} character, as observed also in ref. \cite{DeVita2022} in a photon energy scan across the whole Cr 3\textit{p} absorption edge. Additionally, a weakly dispersing feature at $\sim\SI{-3.1}{\eV}$, as well as two arms of an electron-like dispersive band at the edge of the BZ from \SIrange{-1.5}{-3}{\eV} (highlighted by blue arrows in Fig. \ref{fig:Fig3}a), appear enhanced at resonance. While the former gives a well-defined peak in the corresponding EDC, the intensity of the latter contributes to the whole energy range from \SIrange{-1.5}{-3}{\eV}, therefore no corresponding peak can be identified. Differently from the state at $\sim\SI{-1.3}{\eV}$, our out-of-resonance data do not reveal these other states at all, probably because of unfavorable matrix elements suppressing the photoemitted intensity. The difference in spectral weight of the bright bands at resonance is also likely due to photoemission matrix elements, preventing us from revealing the whole extent of the Cr \textit{d}-orbital manifold in the valence band with uniform intensity.

To facilitate comparison between ARPES measurements and the DFT-calculated spectral function, in Fig.~\ref{fig:Fig3}c we projected the Cr 3\textit{d} bands from the total spectral function shown in Fig.~\ref{fig:Fig1}f. To show the number of electronic states at different energies more transparently, we display the atom-projected density of states (pDOS) side-by-side with the spectral function. From this plot we can observe that the Cr 3\textit{d} and I 5\textit{p} states overlap in a broad range across several electronvolts, suggesting strong orbital mixing between the two. Cr 3\textit{d} states span from the VBM down to \SI{-3.6}{\eV}, displaying two distinct features: larger (broader) one centered around \SI{-1.3}{\eV} and smaller (narrower) one at \SI{-2.8}{\eV} (green highlight in Fig.~\ref{fig:Fig3}d). When the intensity and energy positions of these features are compared to the EDCs acquired at \SI{300}{\K} (Fig.~\ref{fig:Fig3}b), they correspond to regions of high spectral weight arising from the Cr 3\textit{d} bands at \SI{-1.3}{\eV} and \SI{-3.1}{\eV}. Moreover, the distributed intensity between the two local maxima mirrors the presence of the dispersive states from \SIrange{-1.5}{-3}{\eV}. We emphasize that the right choice of Hubbard \textit{U} on Cr 3\textit{d} orbitals is crucial here, as the higher \textit{U} values push down the larger feature too much, so that for $U > \SI{2.0}{\eV}$ the two features merge into a single broad band (Fig.~S7).

\begin{figure}
\includegraphics[width=0.95\linewidth]{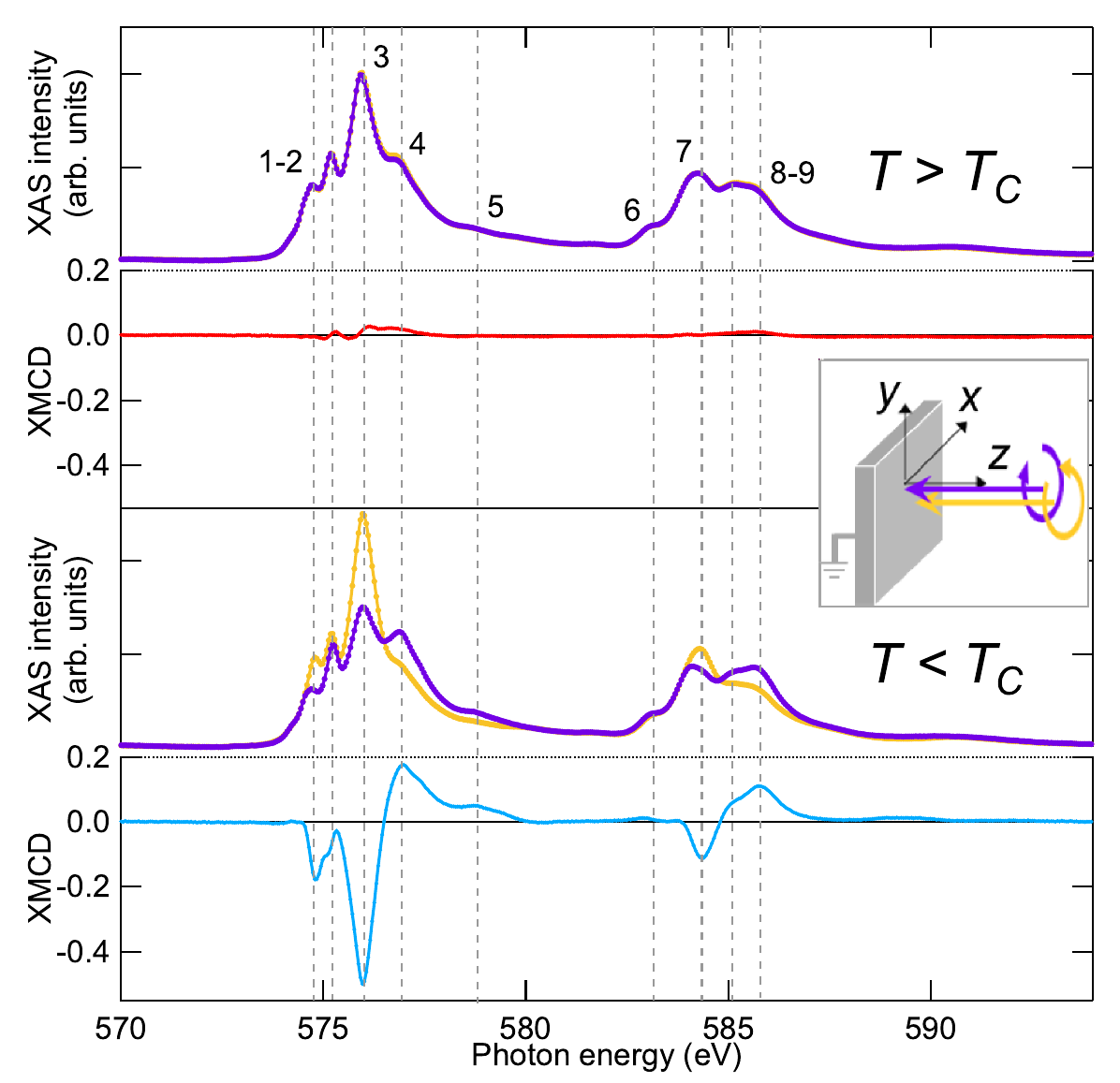}
\centering
\caption{\protect\label{fig:Fig2}XAS spectra across the Cr \textit{L}\textsubscript{2,3} edges, acquired with left and right circularly polarized photons, and corresponding XMCD, at \SI{90}{\K} (\textit{top}) and at \SI{30}{\K} (\textit{bottom}). The inset shows the experimental geometry.}
\end{figure}

Unlike these deeper states that come mostly from Cr 3\textit{d} orbitals, the states just below the VBM are mainly derived from I 5\textit{p} orbitals. This can be rationalized from the disappearance of the prominent dispersing band at the VBM near $\Gamma$ (Fig.~\ref{fig:Fig1}f) in the Cr 3\textit{d}-projected spectral function (Fig.~\ref{fig:Fig3}c). Additionally, the very deep states between $\SI{-3.8}{\eV}$ and $\SI{-4.6}{\eV}$, clearly visible in off-resonance ARPES measurements and suppressed under resonant conditions (Fig~\ref{fig:Fig2}a), exhibit almost exclusively I 5\textit{p} character, which is evident from the pDOS plot in Fig.~\ref{fig:Fig3}d.

Now we turn to the orbital character in the ferromagnetic state of CrI\textsubscript{3}. We present in Fig. \ref{fig:Fig2} the X-ray Absorption Spectroscopy (XAS) and XMCD spectra across the Cr \textit{L}\textsubscript{2,3} edges, below and above $T_C$.
The overall lineshape of the XAS white line \cite{Frisk2018,Kim2019} shows the main \textit{L}\textsubscript{3} peak (3) at \SI{575.9}{\eV}; two features lie at lower (peaks 1,2) and higher (peaks 4,5) photon energies. The \textit{L}\textsubscript{2} peak displays two features giving opposite dichroism (7,9), with a minor peak of the same dichroic character (8) and a pre-edge (6). We also observe almost no difference between $T>T_C$ and $T<T_C$ in the lineshape of the polarization-averaged XAS spectra (Supplemental Material Fig.~S2, main panel), which suggests that sample conditions were extremely stable during the measurement, and that the difference in the XMCDs is fully attributable to intrinsic effects. 

We quantify from sum rules \cite{Chen1995} the orbital and effective spin moments as $m_{L}=\num{0.00\pm0.06} \; \mu_B/\textrm{Cr}$, and $m_{S}=\num{2.45\pm0.05} \; \mu_B/\textrm{Cr}$. The slightly positive measured value of $\mu_L$ is smaller than the error, and compatible with zero orbital moment, as expected in CrI\textsubscript{3}. Conversely, $m_S$ is somewhat lower than the expected value ($3\;\mu_B/\textrm{Cr}$). In any case, the calculation of $m_L$ and especially $m_S$ is not necessarily an accurate procedure for early transition metals, due to the large overlap between \textit{L}\textsubscript{3} and \textit{L}\textsubscript{2} edges \cite{Obrien1994}.

To compare our data, we refer to previous results on CrI\textsubscript{3} \cite{Frisk2018}, as well as the standard reference for pure Cr\textsuperscript{3+} valence states, the oxide Cr\textsubscript{2}O\textsubscript{3} \cite{Ito1999,Kucheyev2007,Chiou2011,Singh2012,Vazquez2018}, similar to CrI\textsubscript{3} in terms of crystal structure, nominal transition metal valence, and ligand coordination geometry. The details of the lineshape and the relative intensity between peaks show clear deviations from ref. \cite{Frisk2018}, which is in turn closer to Cr\textsubscript{2}O\textsubscript{3}. Features located in the low-energy region and in the neighborhood of pre-edges of the \textit{L}\textsubscript{3} and \textit{L}\textsubscript{2} peaks, which are crucial to correctly assess the orbital character in transition metal halides \cite{Sant2023,DeVita2024}, show the most evident differences. In our case, peak (1) is sharper and more pronounced, peak (2) has no equivalent, and peak (6) emerges more clearly from the pre-edge region, other than presenting a non-zero dichroic signal. Lastly, the tail of the \textit{L}\textsubscript{3} peak is considerably less broad. 

We also note that the \textit{L}\textsubscript{2,3} edges appear shifted towards lower photon energies compared to Cr\textsubscript{2}O\textsubscript{3} \cite{Ito1999,Kucheyev2007,Chiou2011,Singh2012,Vazquez2018}, although not strongly  enough to justify an outright change in valence state \cite{Theil1999}. From these observations we gather that the electronic character in CrI\textsubscript{3} shows an excess of electrons compared to a pure Cr\textsuperscript{3+}. A higher electron count at the cation site in CrI\textsubscript{3} does not necessarily imply a mixed valence state: that would affect the lineshape of the XAS white line by considerably broadening spectral features, as seen \textit{e.g.} in mixed-valence manganites \cite{Abbate1992,DeJong2005,Werner2010,Pesquera2016}. The small signal in the pre-edge region around \SI{573}{\eV} (Supplemental Material, Fig.~S2 inset), where typically Cr\textsuperscript{2+} states are located \cite{Theil1999,Ishida2008,Buccoliero2025}, is two orders of magnitude lower compared to the intensity of the main peaks, not nearly enough to support a mixed-valence state. Instead, we interpret our data as a partial redistribution of the charge by the ligands; that is, a more \textit{covalent} character of the bond due to Cr \textit{e\textsubscript{g}} -- I 5\textit{p} orbital hybridization \cite{Liu2018,Shao2021,He2025}. This is further supported by the presence of a dichroic signal of the same sign at the I \textit{M}\textsubscript{4,5} edges (Supplemental Material, Fig.~S3), due to mixing of Cr and I final states \cite{Kim2019}. Moreover, multiplet cluster calculations based on the Quanty code \cite{Haverkort2016} show that, in absence of hybridization, several features at the Cr \textit{L}\textsubscript{2,3} edges appear at odds with the experiment. If hybridization is taken into account, calculations match our experimental lineshape more closely, and we are able to extract a 3\textit{d} electron occupation of $n=3.52$ (Supplemental Material, Fig.~S4 and Table~S1). We attribute the high effective Cr 3\textit{d} occupancy to the strong Cr–I covalency, which defines the local chemical environment and effectively increases the electron count on the Cr site.

\begin{figure*}[htb]
\includegraphics[width=0.9\linewidth]{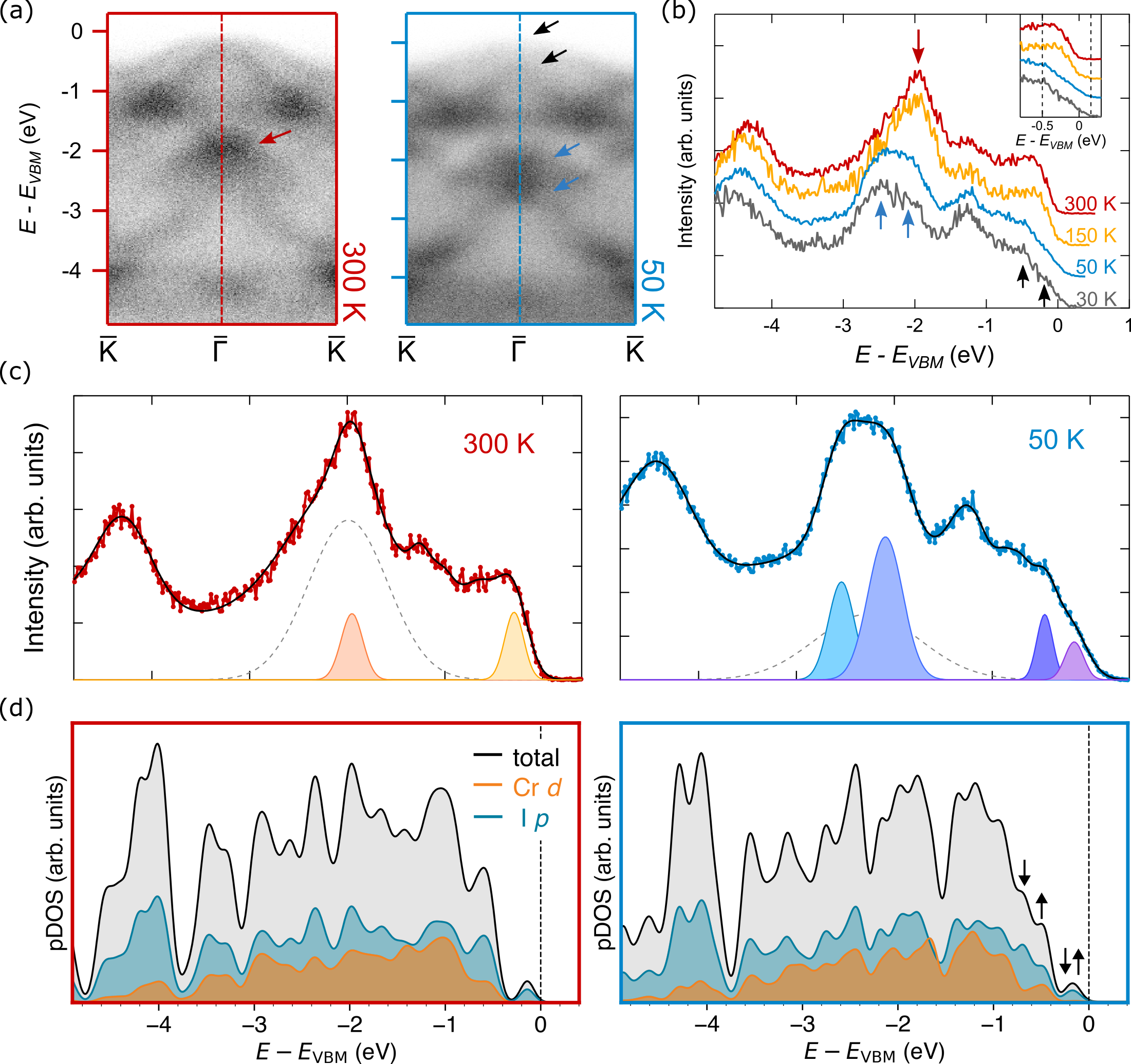}
\caption{\protect\label{fig:Fig4} (a) ARPES (\textit{E}, \textit{k}) spectra along the $\bar{\mathrm{K}}-\bar{\Gamma}-\bar{\mathrm{K}}$ direction, at $T=\SI{300}{\K}$ and $T=\SI{30}{\K}$. (b) EDCs at $\bar{\Gamma}$ as a function of temperature. The inset enhances the VBM region to highlight the change in slope. (c) EDCs at $\bar{\Gamma}$ at \SI{300}{\K} (\textit{left}) and at \SI{50}{\K} (\textit{right}). The black line is a multipeak fit as described in the main text. The colored Gaussians represent the spectral features splitting below $T_C$. (d) Calculated pDOS at the $\Gamma$ point, including \SI{100}{\milli\eV} Gaussian broadening to account for experimental resolution, in the PM (\textit{left}) and FM (\textit{right}) phases projected on Cr \textit{d} (orange) and I \textit{p} (blue) orbitals.}
\end{figure*}

To assess the modifications induced in the electronic structure by the development of the ferromagnetic ordering, we acquired (\textit{E}, \textit{k}) spectra above and below $T_C$, shown in Fig.~\ref{fig:Fig4}a. We extracted EDCs at the $\bar{\Gamma}$ point in Fig. \ref{fig:Fig4}b, to better illustrate the changes occurring in the band dispersion.

By comparing spectra and EDCs, we do not observe significant changes in effective masses within bands, or in the energy position of the features at high BE, indicating that the bandwidth of the VB is not strongly affected. However, two major modifications take place: at $T<T_C$, we can see that (i), the hole-like pocket at $\overline{\Gamma}$ at \SI{-2}{\eV}, corresponding to the most intense peak in the EDCs in Fig. \ref{fig:Fig4}b (red arrow), shifts towards higher BEs and broadens, revealing two separate components (blue arrows); and (ii), the VBM area shows a displacement of the two dispersive I 5\textit{p} bands (black arrows), translating into a noticeable broadening in the EDCs.

When looking at the inset in Fig. \ref{fig:Fig4}b, where the VBM region is zoomed in, the cutoff of the EDC at the VBM is sharper at higher temperatures, contrasting with the expected thermal broadening. Similarly, the feature at \SI{-2}{\eV} broadens at lower temperature, in contrast with the behavior of bands in other regions of the spectrum (Supplemental Material Fig.~S5).

To understand the different spectral contributions, we fitted the EDCs with Gaussian shapes superimposed to a Shirley background, focusing in particular on the region at \SI{-2}{\eV} and the VBM (Fig. \ref{fig:Fig4}c). The EDC at \SI{300}{\kelvin} shows the presence of two overlapping bands, one much broader than the other, marked by different effective masses as evident from the (\textit{E}, \textit{k}) dispersion. On the other hand, at $T=\SI{50}{\K}$ two such peaks appear on top of the broader signal. This is consistent with a splitting of $\approx\SI{450}{\milli\eV}$ of the sharper feature in two separate photoemission signals in the FM state. We notice the same effect at the VBM, where a splitting appears ($\approx\SI{250}{\milli\eV}$). It is also noteworthy that the center of mass of the I 5\textit{p} band manifold at \SI{-2}{\eV} shifts towards lower BEs by $\approx\SI{300}{\milli\eV}$. On the other hand, the shifts of other bands are comparable to our experimental bandwidth ($\approx\SI{100}{\milli\eV}$), therefore no conclusive statement can be made. 

The experimental EDC spectra are further interpreted through the lens of DFT-calculated pDOS evaluated at the $\Gamma$ point. Fig.~\ref{fig:Fig4}d shows how selected single peaks in the PM phase each split into two in the FM phase (small black arrows). The peak at the VBM is of purely I~$5p$ character, whereas the feature situated at approximately \SI{-0.6}{\eV} consists of hybridized spin-majority Cr~$3d$ and I~$5p$ states. In the FM phase, both peaks exhibit an exchange splitting of roughly \SI{100}{\milli\eV}. Given the purely I~$5p$ nature of the VBM, this observation strongly suggests that the ferromagnetic alignment of the Cr magnetic moments, mediated via Cr~$e_g$--I~$5p$ hybridization channels, is highly efficient in splitting the iodine states.

For the deeper states in the EDC spectra -- specifically the single peak near \SI{-2}{\eV} in the PM phase that splits into two features in the FM phase (Fig.~\ref{fig:Fig4}c) -- the interpretation is less straightforward. In this energy range (\SIrange{-1.5}{-3.0}{\eV}), the DFT-calculated pDOS contains several overlapping contributions, which makes a quantitative determination of the splitting difficult.

Symmetry considerations further support that the long-range magnetic ordering in the FM ground state is precisely driven by the orbital mixing. The only symmetry-allowed ligand-to-metal bonds are $\sigma$-type involving \textit{e\textsubscript{g}} and $\pi$-type involving \textit{t\textsubscript{2g}}. When the spin orientation is out-of-plane (\textit{i.e.}, the preferable direction of Cr magnetic moments in bulk CrI\textsubscript{3}), $\pi$ bonding cannot occur between \textit{t\textsubscript{2g}} and spin-orbit entangled $j=1/2$ ligand states \cite{Kim2019}, leaving superexchange $\sigma$ hopping as the preferred path. 
The \textit{e\textsubscript{g}} orbital occupation due to hybridization is certainly low compared to the \textit{t\textsubscript{2g}}; however, as argued by Stavropoulos et al. \cite{Stavropoulos2021}, the interaction given by \textit{e\textsubscript{g}} paths in presence of heavy ligand SOC and strong Hund's coupling is expected to drive the system into a FM state. The upshift of the I 5\textit{p} bands at \SI{-2}{\eV} that we experimentally observe (Fig. \ref{fig:Fig4}c-d) is precisely the signature of Hund's energy gain manifesting on the ligand site, similarly to the mediation of the exchange via Te hybridized bonding in CrGeTe\textsubscript{3} \cite{Watson2020}. Hund's on-site energy is thus minimized, stabilizing parallel spins as the ground-state configuration \cite{Ivanov2023}.  As a result, it is the hybridization between Cr \textit{e\textsubscript{g}} and I \textit{p} orbitals in presence of strong Hund's coupling that allows the stabilization of a FM ground state in CrI\textsubscript{3}.

In any case, we note that the hybridized superexchange picture does not exclude contributions from other mechanisms, \textit{e.g.}, \textit{t\textsubscript{2g}}--\textit{e\textsubscript{g}} hopping, which is also expected to contribute to the FM exchange \cite{Song2022,Ghosh2023}.

\section{Conclusions}

In summary, we presented a combined experimental and theoretical study of the orbital structure of CrI\textsubscript{3}, and its interplay with the emergence of long-range magnetic order in this compound. By comparing (\textit{E}, \textit{k}) non-resonant and resonant ARPES spectra with DFT calculations, we were able to assess the orbital-resolved band ordering. XAS and XMCD measurements at the Cr \textit{L}\textsubscript{2,3} and I \textit{M}\textsubscript{4,5} edges suggest a sizable orbital mixing, with an electron count of \num{3.52} electrons per Cr atom from multiplet cluster calculations. By means of ARPES spectra we reveal multiple large splittings (\SIrange{200}{400}{\milli\eV}) opening in I-derived bands in the FM phase, as well as an explicit band signature (a large $\sim\SI{300}{\milli\eV}$ upshift) of Hund's on-site energy gain. Our results confirm that the ligand-metal covalent orbital character largely supports the onset of the FM phase, and that -- thanks to the orbital mixing -- Hund's coupling drives the charge and magnetic order in CrI\textsubscript{3}, providing the first robust demonstration of the theory put forward in Ref. \cite{Kim2019,Stavropoulos2021} and reinforcing CrI\textsubscript{3} as a Mott-Hund insulator.

The present study focuses on CrI\textsubscript{3} as a prototypical material, but it is realistic to hypothesize that the same principles may be generalized to other vdW magnets. In particular, the considerations on correlations and couplings at an orbital-specific level, framing the electronic/magnetic phase diagram, do not lose generality when extended to other 3\textit{d}-transition-metal-based systems, whenever a source of SOC in the form of a heavy ligand is provided. Given the many structural and chemical/orbital similarities between vdW magnets, we expect that further experimental endeavors may recognize the role of orbital mixing and \textit{d}-orbital electronic correlations in underpinning the magnetic properties of many of these materials.


\section*{Acknowledgments}
This work is financially supported by the Deutsche Forschungsgemeinschaft (DFG) within Transregio TRR 227 Ultrafast Spin Dynamics project B07, the Max Planck Society, and BERLIN QUANTUM, an initiative endowed by the Innovation Promotion Fund of the city of Berlin. We thank the European Synchrotron Radiation Facility (proposal ih-hc-3612 and ih-hc-3640) and the Elettra Synchrotron for provision of beamtime; we also thank D. Betto for assistance in using beamline ID32, and J. Fujii for assistance in using beamline APE-LE. G.P. acknowledges support from Progetto PRIN 2022 “BEST-WIN” PNRR Next Generation EU – CUP B53D23004530006, and from PNRR-MUR project PE0000023-NQSTI. S.P. acknowledge partial financial support by the Next-Generation-EU program via the PRIN-2022 SORBET (Grant No. 2022ZY8HJY). S.S. acknowledges financial support from the Vin\v{c}a Institute, provided by the Ministry of Science, Technological Development and Innovation of the Republic of Serbia through the contract No. 451-03-33/2026-03/200017. Computational resources and support were provided by CINECA under the ISCRA initiative, through the projects ISCRA-B HP10BA00W3 and ISCRA-C HP10C6WZ1O. S.S. and S.P. acknowledge support from the Ministry of Foreign Affairs of Italy and the Ministry of Science, Technological Development, and Innovation of Serbia through the bilateral project ``Van der Waals Heterostructures for Altermagnetic Spintronics'', realized under the executive programme for scientific and technological cooperation between the two countries. This work was partially performed in the framework of the NFFA-SPRINT facility, supported by MUR as the Activity of International Relevance NFFA, (www.trieste.NFFA.eu). 
The authors gratefully thank Dr. P. Stavropoulos and Dr. F. Barantani for valuable insight and fruitful discussion.

\section*{Author Contributions}

T.P., A.D.V, and R.E. conceived the research project. A.D.V., T.P., and R.S. performed the experiments. N.B.B. and I.V. provided and supported access to synchrotron facilities. A.D.V. analyzed the experimental data. DFT calculations were conducted by S.S. under the supervision of S.P. R.S. developed the code and performed multiplet cluster calculations. L.R. participated in maintaining and running the experimental apparatus. R.E., M.W., and G.P. provided funding and scientific supervision for the experiments. A.D.V. and S.S. wrote the manuscript, with contributions from all co-authors. 

\section*{Data availability}

The data that support the findings of this article are openly available \cite{DataAvailability}.

\section*{Appendix: DFT methods}

\noindent\textbf{DFT modeling of the paramagnetic state} -- The magnetic anisotropy in a CrI$_3$ monolayer favors the out-of-plane magnetization direction. Therefore, we modeled the FM state with Cr magnetic moments of $3 \, \mu_{\rm B}$ aligned out-of-plane and parallel to each other.
On the other hand, modeling the PM state is not trivial within DFT approach. For this cause we used the \textit{special quasi-random structures} (SQS) approach for binary alloys~\cite{Zunger1990Jul} where instead of having two atomic types A and B like in binary alloys, we have two spin types distributed over a honeycomb lattice of Cr atoms. In particular, we constructed a $3 \times 3 \times 1$ supercell and used the {\sc icet} tool~\cite{Angqvist2019Jul} to distribute 9 spin-up and 9 spin-down moments in a quasi-random fashion over 18 Cr atoms. SQS structures optimally approximate random alloys by matching the correlation functions of truly random systems up to the $n$-th nearest-neighbor shell, providing the closest structural representation for computational modeling of disordered solid solutions. This approach ensures that  CrI\textsubscript{3} has zero net magnetization while suppressing long-range magnetic order. Note that similar approach was used in Ref.~\cite{Chen2020Dec} to model the PM state in the same compound.
It should be noted that DFT is a zero-temperature theory, which inherently limits its ability to fully describe the physical properties of the magnetic system in the real PM phase at higher temperatures. In this regard, the DFT-simulated PM state represents an approximation where Cr magnetic moments retain their magnitude ($3 \, \mu_{\rm B}$), while the whole structure exhibits zero net magnetization and lacks long-range magnetic order.\par

\noindent\textbf{Calculation of the spectral function} -- The spectral function, calculated with the supercell, is unfolded to the unit cell to facilitate comparison with ARPES data. The unfolding procedure is performed using the post-processing package {\sc bands4vasp} \cite{Dirnberger2021Jun} where the spectral function is defined as $A(\mathbf{k},E) = \sum_m P_{\mathbf{K}m}(\mathbf{k}) \delta(E-E_m)$, where $\mathbf{k}$ $(\mathbf{K})$ is the wavevector in the BZ of the unit cell (supercell) and $P_{\mathbf{K} m}(\mathbf{k})$ projects the supercell eigenstates $\ket{{\mathbf{K}} m}$ onto the unit cell eigenstates $\ket{\mathbf{k} n}$. By calculating the spectral function rather than just the band structure, we can identify which $\mathbf{k}$-resolved states vanish due to magnetic disorder and which remain robust. This approach directly reveals how magnetic disorder affects the electronic structure in different energy ranges and different regions of the BZ.

\noindent\textbf{The choice of Hubbard and Hund parameters} -- In the work of Haddadi \textit{et al.}\cite{Haddadi2024Jan}, it is demonstrated that PBEsol xc functional provides a closer-to-experiments description of the electronic bands of CrI\textsubscript{3} compared to the PBE functional. We note that some kind of on-site energy correction is expected in CrI$_3$, as the insulating ground state, despite the partially filled Cr 3\textit{d} orbital, indicates a Mott-Hubbard-driven stabilization of a bandgap \cite{Webster2018}. Therefore, we adopted the DFT+\textit{U} approach as introduced by Liechtenstein \textit{et al.} \cite{Liechtenstein1995} and tested several combinations of $(U,J)$ values, with $U$ ranging from $1 - 3 \, {\rm eV}$ and $J$ from $0.6 - 0.9 \, {\rm eV}$. Spectral function plots for different $(U,J)$ values are presented in Fig.~S5. As the main criterion, we examined how the most prominent Cr $3d$ bands calculated with DFT$+U$ in the PM state reproduce those measured with ARPES at $300 \, {\rm K}$. Ultimately, we decided to use Hubbard $U = 1.3 \, {\rm eV}$ and Hund $J = 0.9 \, {\rm eV}$. This particular set of $(U,J)$ values, besides providing good overall fit of ARPES data, allows us to strike a balance between the findings of Stavropoulos \textit{et al.} \cite{Stavropoulos2021}, who argue that a finite Hund coupling $J$ is essential for stabilizing ferromagnetic order in CrI\textsubscript{3}, and the results of Haddadi \textit{et al.} \cite{Haddadi2024Jan}, which demonstrate that the PBEsol functional with a small Hubbard $U$ provides good agreement overall with experimental electronic structure data. As argued in Ref.~\cite{Haddadi2024Jan}, the agreement between DFT-calculated band structure and ARPES data further improves when two distinct Hubbard \textit{U} parameters are used for the spin-up and spin-down channels, though we ultimately decided to use a single \textit{U} for both channels. We also note that varying $J$ in the range $0.6 - 0.9 \, {\rm eV}$ has minimal effect on the calculated spectral function.

\bibliography{bibliography}

\end{document}



\title{Supplemental Material: Orbital mixing and strong Hund's coupling stabilize spin order in van der Waals ferromagnet CrI\textsubscript{3}}

\author{A. De Vita}
\email{alessandro.de.vita@tu-berlin.de}%
\affiliation{Institut für Physik und Astronomie, Technische Universität Berlin, Straße des 17 Juni 135, 10623 Berlin, Germany\looseness=-1}%
\affiliation{Fritz Haber Institute of the Max Planck Society, Faradayweg 4--6, 14195 Berlin, Germany\looseness=-1}%
\author{S. Stavri\'c}
\email{stavric@vin.bg.ac.rs}%
\affiliation{Vin\v{c}a Institute of Nuclear Sciences - National Institute of the Republic of Serbia, University of Belgrade, P. O. Box 522, RS-11001 Belgrade, Serbia\looseness=-1}%
\author{R. Sant}%
\affiliation{ESRF, The European Synchrotron, 71 Avenue des Martyrs, CS40220, 38043 Grenoble Cedex 9, France\looseness=-1}%
\affiliation{Dipartimento di Fisica, Politecnico di Milano, Piazza Leonardo da Vinci 32, I-20133 Milano, Italy\looseness=-1}%
\author{N. B. Brookes}%
\affiliation{ESRF, The European Synchrotron, 71 Avenue des Martyrs, CS40220, 38043 Grenoble Cedex 9, France\looseness=-1}%
\author{I. Vobornik}
\affiliation{Consiglio Nazionale delle Ricerche CNR-IOM, Unità di Trieste, Strada Statale 14, km 163.5, 34149 Basovizza (TS), Italy\looseness=-1}%
\author{G. Panaccione}
\affiliation{Consiglio Nazionale delle Ricerche CNR-IOM, Unità di Trieste, Strada Statale 14, km 163.5, 34149 Basovizza (TS), Italy\looseness=-1}%
\author{S. Picozzi}
\affiliation{Department of Materials Science, University of Milan-Bicocca, Via Roberto Cozzi 55, 20125 Milan, Italy\looseness=-1}
\affiliation{Consiglio Nazionale delle Ricerche CNR-SPIN, c/o Università degli Studi ‘G. D’Annunzio’, 66100 Chieti, Italy\looseness=-1}%
\author{M. Wolf}
\affiliation{Fritz Haber Institute of the Max Planck Society, Faradayweg 4--6, 14195 Berlin, Germany\looseness=-1}%
\author{L. Rettig}
\affiliation{Fritz Haber Institute of the Max Planck Society, Faradayweg 4--6, 14195 Berlin, Germany\looseness=-1}%
\author{R. Ernstorfer}
\email{ernstorfer@tu-berlin.de}%
\affiliation{Institut für Physik und Astronomie, Technische Universität Berlin, Straße des 17 Juni 135, 10623 Berlin, Germany\looseness=-1}%
\affiliation{Fritz Haber Institute of the Max Planck Society, Faradayweg 4--6, 14195 Berlin, Germany\looseness=-1}%
\author{T. Pincelli}
\email{pincelli@tu-berlin.de}%
\affiliation{Institut für Physik und Astronomie, Technische Universität Berlin, Straße des 17 Juni 135, 10623 Berlin, Germany\looseness=-1}%
\affiliation{Fritz Haber Institute of the Max Planck Society, Faradayweg 4--6, 14195 Berlin, Germany\looseness=-1}%

\maketitle

\section*{Additional Figures}

For completeness, we present in Fig.~\ref{fig:SIle} an example of the extrapolation of the leading edge to calibrate the zero of the $E-E_{VBM}$ scale in ARPES measurements.

In Fig.~\ref{fig:SI1} we display the average between left and right circularly polarized photons of the XAS signal across the Cr \textit{L}\textsubscript{2,3} edges, above and below $T_C$, to show the absence of significant modifications of the lineshape with the change in temperature. The inset, zooming in the pre-edge region, emphasizes the presence of an additional spectroscopic signal with an intensity two orders of magnitude smaller than the main peak, as discussed in the main text.

\renewcommand{\thefigure}{S1}
\begin{figure}[htb]
\centering
\includegraphics[width=0.65\linewidth]{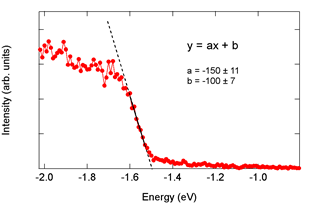}
\caption[captionsep=none]{\protect\label{fig:SIle} EDC of CrI\textsubscript{3} at \SI{50}{\K}; the dashed line represents the fit of the leading edge, whose parameters are outlined in the inset.}
\end{figure}

\renewcommand{\thefigure}{S2}
\begin{figure}[htb]
\centering
\includegraphics[width=0.45\linewidth]{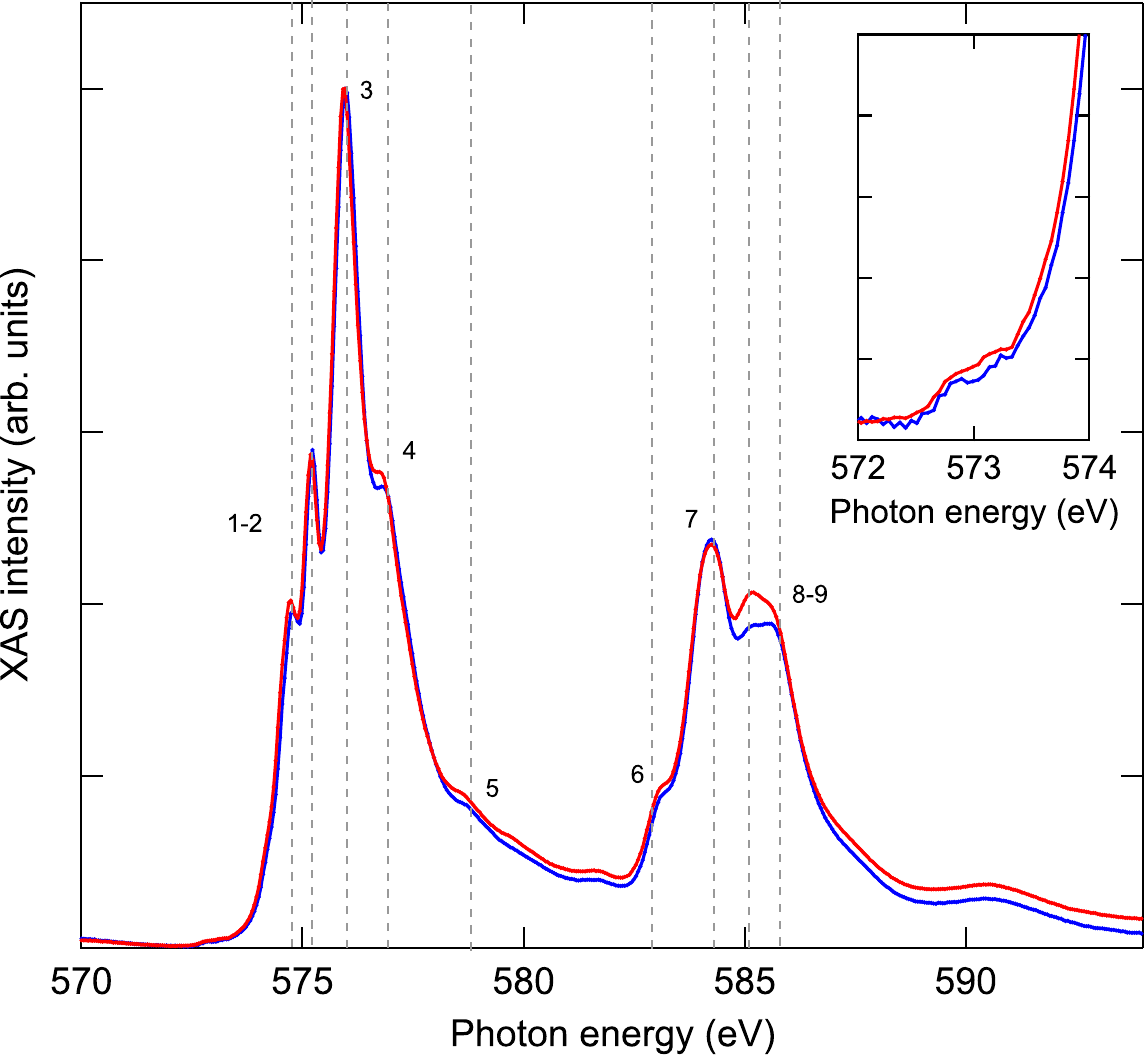}
\caption[captionsep=none]{\protect\label{fig:SI1} XAS spectra across the Cr \textit{L}\textsubscript{2,3} edges, averaged between left and right circularly polarized photons, acquired at \SI{30}{\K} (blue) and at \SI{90}{\K} (red). \textit{Inset}: zoom in the pre-edge region of the same spectra.}
\end{figure}

Additionally, we also present the XAS and XMCD spectra across the I \textit{M}\textsubscript{4,5} edges in \ref{fig:SI3}, highlighting the presence of a non-zero dichroism. The background in the XAS spectra, increasing at higher photon energy, does not correspond to other absorption edges, and is probably a charging effect taking place upon absorption at low temperatures.

\renewcommand{\thefigure}{S3}
\begin{figure}[htb]
\centering
\includegraphics[width=0.45\linewidth]{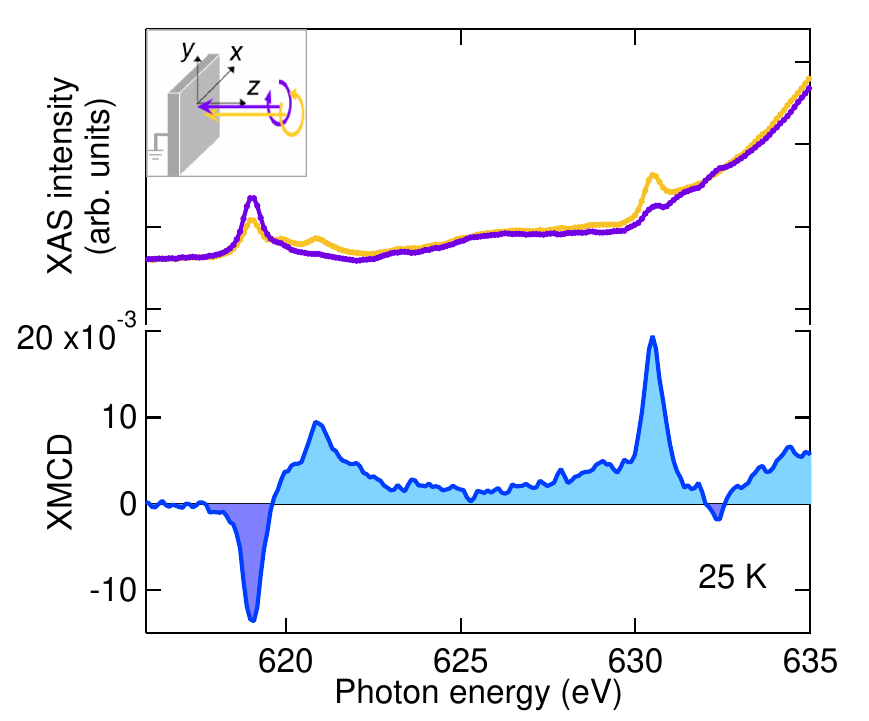}
\caption{\protect\label{fig:SI3} XAS spectra across the I \textit{M}\textsubscript{4,5} edges, acquired with left and right circularly polarized photons, and corresponding XMCD, at \SI{25}{\K}. The inset shows the experimental geometry.}
\end{figure}

In Fig. \ref{fig:SI4} we show cluster calculations with and without hybridization. Black arrows highlight the spectral features most indicative of a hybridized state by comparison with the experiment. The dotted grey line marks the position of the shoulder on the main \textit{L}\textsubscript{3} peak, appearing only in the case without hybridization and not appearing in the experiment at all. In table \ref{tab:SItab1} the parameters for the calculations are presented. $V_{t2g}$ and $V_{eg}$ are the hybridization parameters for each of the given O\textsubscript{h} symmetry sub-groups, expressed in \si{\eV}; $N_{t2g}$, $N_{eg}$ and $N_{3d,\mathrm{tot}}$ are the corresponding electron counts and the total \textit{d}-count for the whole orbital, expressed in e\textsuperscript{-}/atom.

\renewcommand{\thefigure}{S4}
\begin{figure}[htb]
\centering
\includegraphics[width=0.85\linewidth]{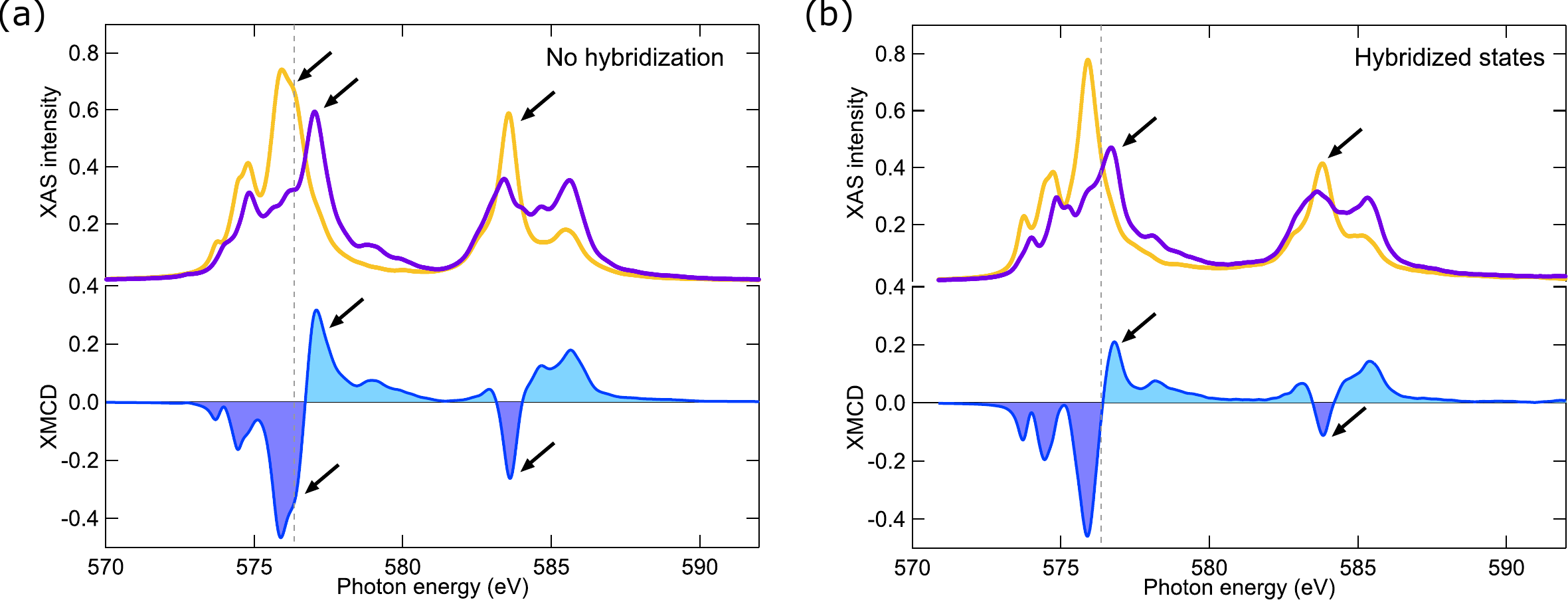}
\caption{\protect\label{fig:SI4} Effects of hybridization on XAS and XMCD spectra. (a-b) Calculated XAS and XMCD spectra at 20 K and normal incidence, without (a) and with (b) hybridization parameters. Left- and right-circularly polarized data are shown with yellow and violet colors respectively. Black arrows highlight the most striking differences of the experiment with the \textit{L}\textsubscript{3} and \textit{L}\textsubscript{2} peaks, respectively. The dashed grey line marks the energy position of the shoulder on the \textit{L}\textsubscript{3} peak.}
\end{figure}

\renewcommand{\thetable}{S1}
\begin{table}[htb]
    \centering
    \begin{tabular}{cSSSSS}
    \toprule
         & {$V_{t2g}$} & {$V_{eg}$} & {$N_{t2g}$} & {$N_{eg}$} & {$N_{3d,\mathrm{tot}}$} \\ 
    \midrule
       Hybrid.  & 1.4 & 1.85 & 3.17 & 0.35 & 3.52 \\ 
       No hybrid.  & 0 & 0 & 3 & 0 & 3 \\ 
    \bottomrule
    \end{tabular}
    \caption{Values of the hybridization parameters and electron count for the cluster calculation in Fig. \ref{fig:SI4}.}
    \label{tab:SItab1}
\end{table}

In Fig. \ref{fig:SI2} we show two EDCs taken at $k_x=\SI{0.55}{\angstrom^{-1}}$ from (\textit{E}, \textit{k}) ARPES spectra above and below $T_C$. The curves show several peaks -- highlighted with black arrows -- getting more defined with decreasing temperature, as expected for electronic states, and differently from the features at the $\bar{\Gamma}$ point, discussed in the main text.

\renewcommand{\thefigure}{S5}
\begin{figure}[htb]
\centering
\includegraphics[width=0.45\linewidth]{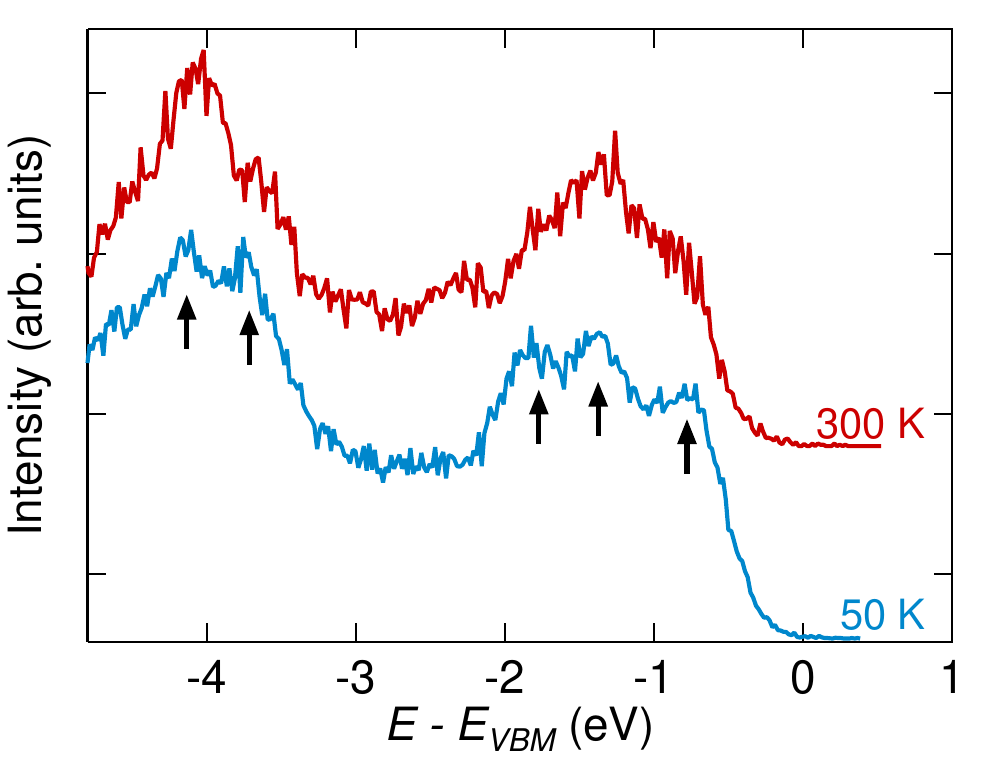}
\caption{\protect\label{fig:SI2} EDCs at \SI{0.55}{\angstrom^{-1}} at \SI{300}{\K} (red line) and at \SI{50}{\K} (blue line). Black arrows highlight the main peaks getting sharper with decreasing temperature.}
\end{figure}

\renewcommand{\thefigure}{S6}
\begin{figure}[htb]
\centering
\includegraphics[width=0.75\linewidth]{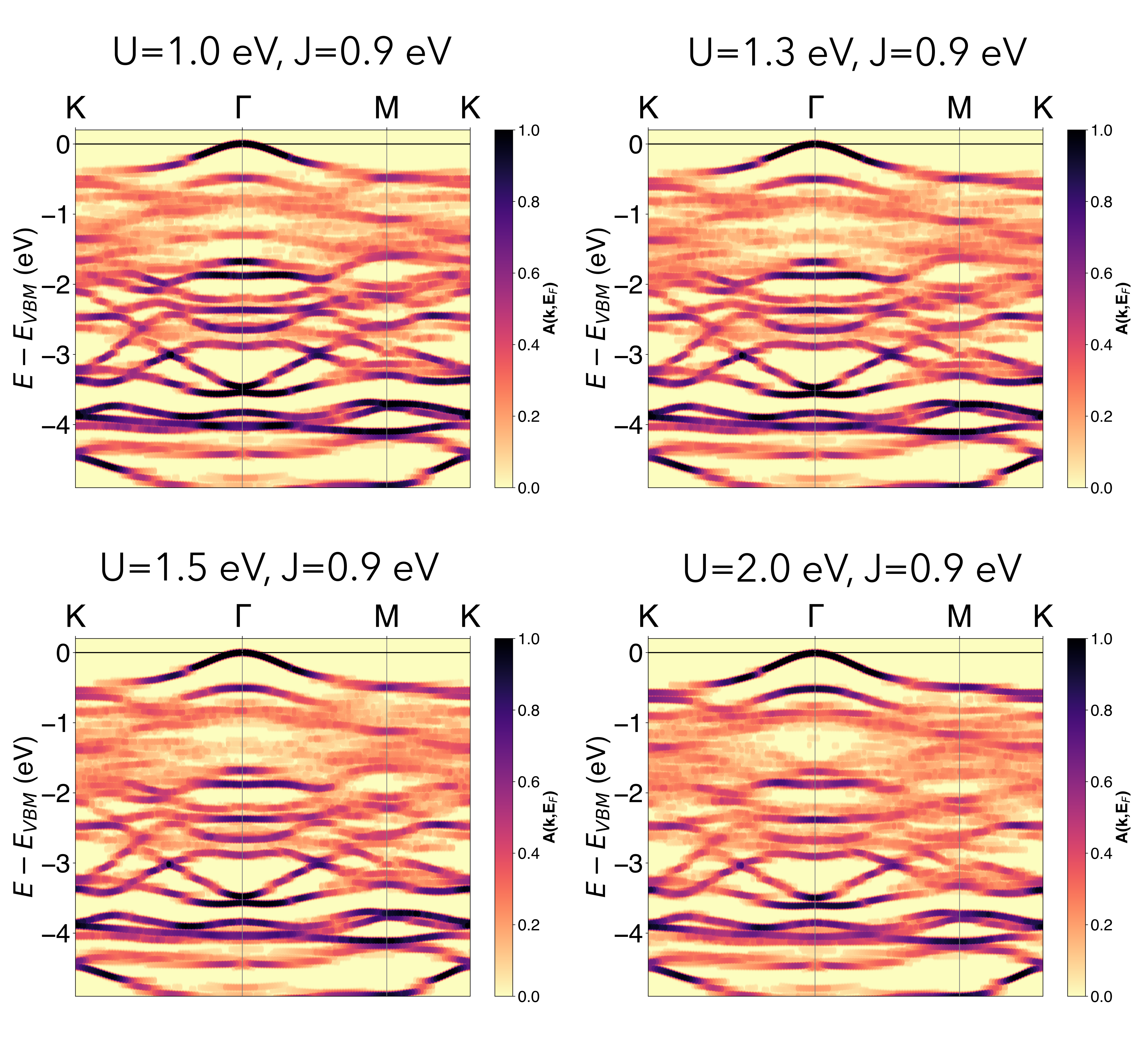}
\caption{\protect\label{fig:SI5} Spectral function of CrI\textsubscript{3} monolayer calculated in the paramagnetic state within DFT+\textit{U} approach 
for different (\textit{U}, \textit{J}) values.}
\end{figure}

\renewcommand{\thefigure}{S7}
\begin{figure}[htb]
\centering
\includegraphics[width=0.75\linewidth]{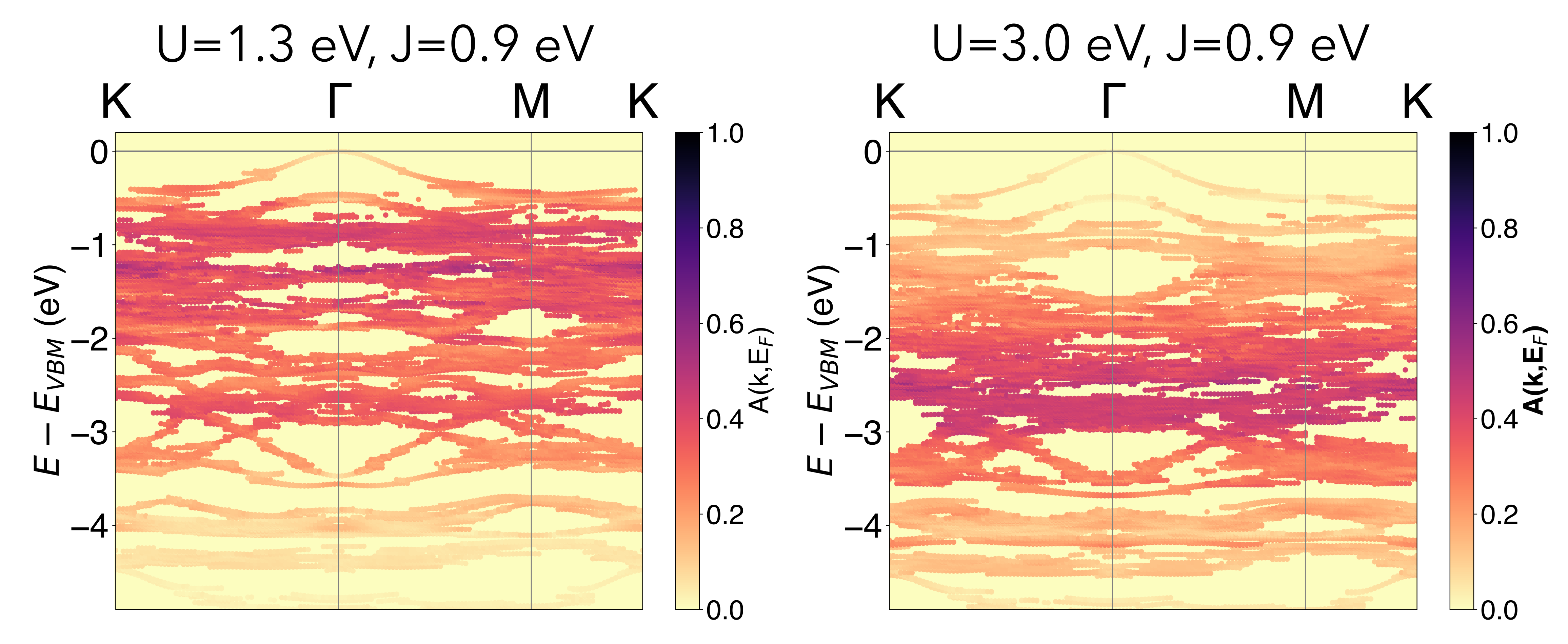}
\caption{\protect\label{fig:SI6} Spectral function of monolayer CrI\textsubscript{3} projected onto Cr-\textit{d} states, computed for two parameter sets: (left) $U=\SI{1.3}{\eV}$, $J=\SI{0.9}{\eV}$ (values adopted in this work) and (right) $U=\SI{3.0}{\eV}$, $J=\SI{0.9}{\eV}$. The left panel reveals two distinct maxima (a stronger feature at \SI{-1.3}{\eV} and a weaker one at \SI{-2.8}{\eV}), while the right panel demonstrates how increasing \textit{U} merges these into a single peak centered at \SI{-2.5}{\eV}.}
\end{figure}

Fig.~\ref{fig:SI5} displays the dependence of the spectral function of CrI\textsubscript{3} on the Hubbard \textit{U}, and Fig.~\ref{fig:SI6} shows the projection of the spectral function onto Cr-\textit{d} states for two \textit{U} values.

